# Using Exclusive Web Crawlers to Store Better Results in Search Engines' Database

Ali Tourani[1] and Amir Seyed Danesh[2]

[1]Department of Software Engineering, University of Guilan, Iran
`a.tourani1991@gmail.com`

[2]Department of Software Engineering, Faculty of Computer Science and Information Technology University of Malaya,
50603, Kuala Lumpur
`Amir.s.d@gmail.com`

## Abstract

*Crawler-based search engines are the mostly used search engines among web and Internet users , involve web crawling, storing in database, ranking, indexing and displaying to the user. But it is noteworthy that because of increasing changes in web sites search engines suffer high time and transfers costs which are consumed to investigate the existence of each page in database while crawling, updating database and even investigating its existence in any crawling operations.*

*"Exclusive Web Crawler" proposes guidelines for crawling features, links, media and other elements and to store crawling results in a certain table in its database on the web. With doing this, search engines store each site's tables in their databases and implement their ranking results on them. Thus, accuracy of data in every table (and its being up-to-date) is ensured and no 404 result is shown in search results since, in fact, this data crawler crawls data entered by webmaster and the database stores whatever he wants to display.*

## 1.Introduction

Searching is a daily activity of many people. Individuals deal with this task every day. The need to search enhances along with the increasing demand for more knowledge and information.

Also, in computer science search and communication are among most important items so that they can be seen in most applications, programs and even on the web and Internet. Nowadays the Internet contains millions of information containing pages. Knowing a site's address users can connect it and use its resources and information. In order to reach this huge information resource human has began to increasingly use search engines. For this reason search engines repeat their crawling, ranking and indexing operations every day. [12]

Among these daily operations crawling specifies the most basic function of a search engine. The biggest search engines of the world do this operation in offline mode (query is not accomplished while entering but it is already done) [16] and its optimization does not influence response received by the user but better results can be embedded in the database through optimization. [17] Making webmasters responsible for storing operations through **"exclusive crawlers"** it can be expected that better results are observed in output sent to the user by filling the search engine's database with a database which contains results of exclusive web-based crawling. Exclusive crawler can act as an exe file, a web site, a php code or a Toolbar for webmasters.

## 1.Architecture of Common Web Crawlers

Main functions of a crawler include downloading web pages, extracting links of each page and following those links. It is also used in web page information processing such as page data mining and email address caning by spammers. [4] But what we consider here is application of crawlers in crawler-based search engines.

A crawler can be divided into two parts: the Spider and the Crawl Manager. [1]
Spider (or Crawler or Robot) is a software the function of which is to search and collect data required by the search engines. [18] The Spider checks different pages, reads their contents, downloads them, follows links, stores and saves required data and finally makes them accessible for other parts of the search engine. [2] Another important part of a Crawler is the Crawl Manager which software (like the Spider) but it manages and commands the Spider. It specifies pages to be reviewed by the Spider. Each of these parts is explained below:

**1.1. The Spider:** In fact a Spider or a Robot is a program (in a common programming languages) responsible for the most primary function. The program is composed of various parts which generally quest web pages and store them in the database, then go to other pages and repeat the procedure.[13] The Spider must quest all available links as web pages contain links to other pages, whether an internal link to their own page or an external link to other pages. [3] If a link is already quested, it will not be quested again.[4][3] Therefore, there is the need to have a data structure that checks this condition. This data structure is irrelevant to the present paper, thus it is not discussed here.

Spider designers optimize the Spider by optimization methods and using time analysis algorithms so that it crawls in the least time with the highest efficiency. There are various optimization algorithms available among which are the Genetic Algorithm [9] and Spreading Activation [10][11] Algorithm. Implementation of each algorithm has certain characterizations relative to the Spider's function. In general, the overall form of a a Spider is as <u>Figure 1.</u>

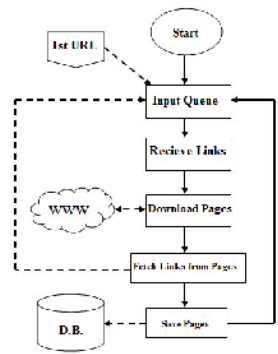

Figure 1. The general structure of Spiders

**1.2. Manager:** In fact, the main role of the Manager is to manage and command Spiders. Obviously a Spider can not quest the whole web. Therefore, each search engines needs several (perhaps tens, hundreds or even thousands) Spiders. Moreover, a searcher can be made in a way in which every Spider crawls a certain part of the web. The more Spiders are recruited, the more up-to-date information is expected to enter the database. [4][21]

The Manager must be implemented in a manner that does not interrupt Spiders. Also, it has to divide a general list to several smaller ones based on which a certain number of pages are being quested in a particular time period. On the other hand, the Manager is responsible for time categorization of Spiders it has to specify conditions.

Each Spider has an exclusive and specific queue which differs from that of the Manager. The Spider, each time, picks an element from Manager's queue and puts that in its own queue. If there is a link in the web page being quested, it is added to the end of the queue. Hence, it is likely for Manager's queue to be composed of hundred and even thousands of elements. Figure2 displays its structure.

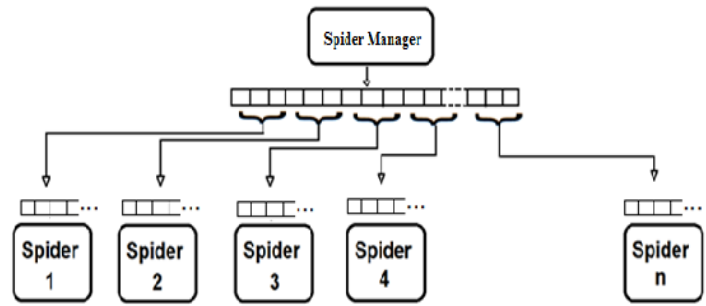

Figure 2. The general structure of a crawler (spider) and its divisions

## 2.2. General Objectives of a Spider Manager

Now we need and algorithm to make Spiders concurrent in the best and most optimized manner. The planned algorithm must possess some certain features. It must:

1) Define the function of each Robot.
2) Determine the time interval to crawl a certain web page.
3) Specify crawling operations strategy. Some crawlers determine the strategy based on page importance and some others based on page changes. [8]
4) Divide entering queue in way that each proper number of links being allocated to each Robot. In fact, every Robot takes an appropriate share of the web.
5) In order to be optimized there must be no overlap in Robots' operation. That is, a link must be checked by several Robots. To meet this need the quest of each links should be checked with a certain ID in the database. [5]
6) Spiders must run simultaneously through "multi-threading programming". Web crawler is usually composed of some threads which work concurrently. Every thread has a specific function and is planned for certain addresses. Hence, more pages are being searched and searching time reduces. "Surface Tree Crawling" is the best method for this goal. Multi-threading must be accomplished in a way that the host does not encounter an overload as a result of demand abundance. [7]

## 3. Problems facing Spiders and crawling operations

Obviously, Spiders face a huge mass of problems while questing the large-scale information on the web. One of these problems is the huge and increasing number of web pages. This rapid growth in the number of web pages makes search engines in trouble in updating information in their databases.[14] Since there is higher demand for crawling operation and a sensible increase is observed in the number of Spiders. It is always likely for some pages to be not crawled or a web site's changes are not presented to users. [20][19]

## 4. Web Exclusive Spider

Exclusive crawler (spider) is a program written in a programming language which is implemented by webmasters (and not search engines) makes responsible for their own site's crawling functions. In fact, it averts the huge mass of processes from search engines and gives webmasters small and insignificant responsibilities.

The exclusive crawler possesses a simpler structure in comparison with search engine crawlers. As mentioned earlier, search engine crawlers are responsible for questing a web page. If there is a link from site A to site B, then the site B must also be quested. But the exclusive does not quest the site found in links and just crawls the host. The comparison between these two procedures is shown in <u>Figure 3.</u>

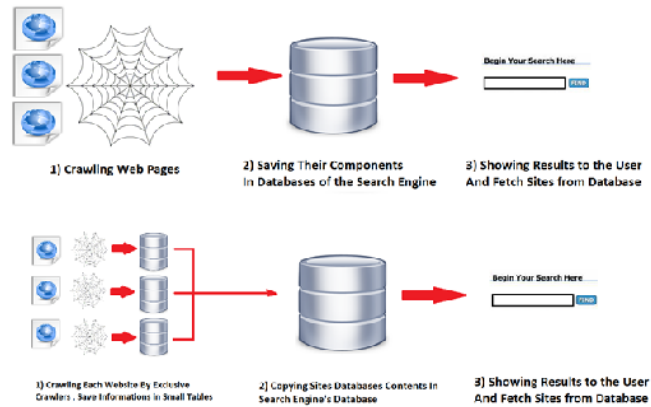

Figure 3. Upper Picture Shows The overal structure of current search engines And the Second Shows The proposed structure using web-based exclusive crawler.

## 4.1. Designing an Exclusive Spider and its Implementation

Like search engine spiders, exclusive spiders are also composed of some basic functions but there are some differences which are explained:

**4.1.1- Input Function:** this function receives the entering queue of links and quests them sequentially. In fact, they receive every element in the list (which is a link) as an entry. The first field in this list includes the first page of the website. Next fields are filled with various pages of the same site. Then, it connects to the Internet, goes to the address and crawls its components (HTML tags). The most basic and fundamental responsibility of a Spider is summarized in this function. Its pseudo code is as follows:

*While (List Is Not Empty)*
*Get First Member of the list which is Homepage of the Website*
*Delete it from List*
*Get the components of URL*
*Go To Parser Function ()*
*Go To Follower Function ()*
*Back to While*

It must be noted that every web site has a part called "robots.text" (to access to this part one can add /robots.text to the end of web site address). This part is identified by Robots' output standard or protocol name and is specified by administrators. This protocol is search engine specific and allows access to the considered web site. [1] There is a function in the structure of current crawlers as checker function. It receives the link from input and enters Robots page and goes to the next step if search permission is available. Otherwise, it can not quest the site. This permission is embodied in the "User-Agent" part of Robots page. [15] If a site does not permit a Robot to access, Spider should not check it and has to receive another input from the queue.

Since the exclusive Spider in single-purpose and is designed for its won host site, there is no need to a function with the same performance. If the webmaster wants some parts to be excluded from final results of a search engine (and not saved in the database) he should not crawl that part.

**4.1.2- Parser Function:** Crawling a page by the Input activates this function. It downloads page components and takes them to various parts. This parsing is partly dependent on Spider architecture and database need. It usually checks tags and parses parts required and important for searching, such as Title and Keywords tags. One needs regular terms with Tree structure to accomplish parsing.

*Get the Component of Page from Input Function*
*Parse Meta-tags of Page to Different Parts such as Title, Keyword, Description, Image Tag, etc...*
*Go To Saver Function ()*

**4.1.3- Saver Function:** This function saves outputs of the Parser in the database. Based on the design and architecture of the database keywords, titles, tag links, page components, image tags, etc. are stored in Tables available in the database. In the implemented exclusive crawler Title, Keyword and Description tags (as well as page components) are put in a certain database.

There are some configurations to optimize saving the whole page by which the Spider only saves the first 1/3 of the site in the database since most search terms are traceable in them. [6]

*Get Different Parts from Parser Function ()*
*Save Them into Database*

 For instance, a part of a database We have designed includes following fields and my exclusive crawler saves pages in its Parser with the order in **Figure 4.**

| ID | Page Number | Title | Description | Keyword | Page Component | Images | Links to | ...... |
|----|-------------|-------|-------------|---------|----------------|--------|----------|--------|

Figure 4. Some of  My Table Columns

**4.1.4- Follower Function:** The function identifies internal links in page (if any) and adds them to the end of input list to be crawled. Therefore, the input queue becomes longer every moment to the extent that all links are being checked. Hence, there is no need to have conditions to avoid overload.

But if there is a link to other web sites, it must not be crawled in the exclusive crawler. In fact, the exclusive crawler is not responsible for it. Thus, we expect a more rapid crawling than search engine crawlers since they lack such a limitation.

*Check the Component of the Page*
*If There Were "Link" Tags push them Back in the Spider List*
*While (the Spider List is not empty)*
*For each URLs in that list*
*Parse them and get Title , Keyword , etc by Parser Function()*
*Save Them Into Database By Saver Function()*
*Else Get back to Input Function ()*

The overall algorithm of the implemented Spider is as **Figure 5.**

```
Void Input () {
    Define Queue[n]
    Save homepage url in first element of Queue
    While ( Queue is not empty ){
      Crawl the page
      Parser(){
       Parse the page to diffrent parts like "keyword","title",etc
      Saver(){
        Save each element in its own field in site Database
       }
      Follower(){
        if there were inside links to diffrent parts of website add them at the end of Queue
        if there were links to other sites do nothing
       }
    }
 }
}
```

Figure 5.  Pseudo code of an exclusive crawler

## 4.2. Exclusive Crawl Manager

Similar to search engine Spider managers having command to coordinate, control and make concurrent Spiders it is possible to design an exclusive spider which manages and command several exclisuve Spiders. This is done for big sites in order to facilitate crawlinf procedure since it is time consuming to crawl big sites with a single Spider.

## 5. An overview of Exclusive Crawler

Using exclsuive crawlers significantly reduces search engine servers' overload in crawling operation. Although application of search engine crawlers (to crawl big sites such as Wikipedia or old sites not updated for a long time) is observed, but this does not meet active webmasters' expectations of information update in search engine database and show the results to users. The webmaster must spend a long time on crawling by exclusive crawler but the updated information is seen in the search engine database shortly after update process is finished.

The exclusive crawler can run as an exe file, a php code or a toolbar for the webmaster. Whenever he updates his site he can crawl once to put updated data in the database (obviously this database differs from those in which memebrs' infromation and specifications are held. In fact, crawling data is saved in a table in the section of site's files).

For example, my exclusive crawler which is written in C++ language crawls each page in an average period of 1.6 seconds (depending on the page size). Conditions of my experiment include: a computer having a 1.73 Corei7 Processor, Cache 6MB, RAM 4GB in which both normal and exclusive crawler programs are written in C++ language without .NET Framework (Desktop form). Besides, they both have a Manager written in Java which runs 10 crawlers concurrently. Results are shown in the **Figure 6**:

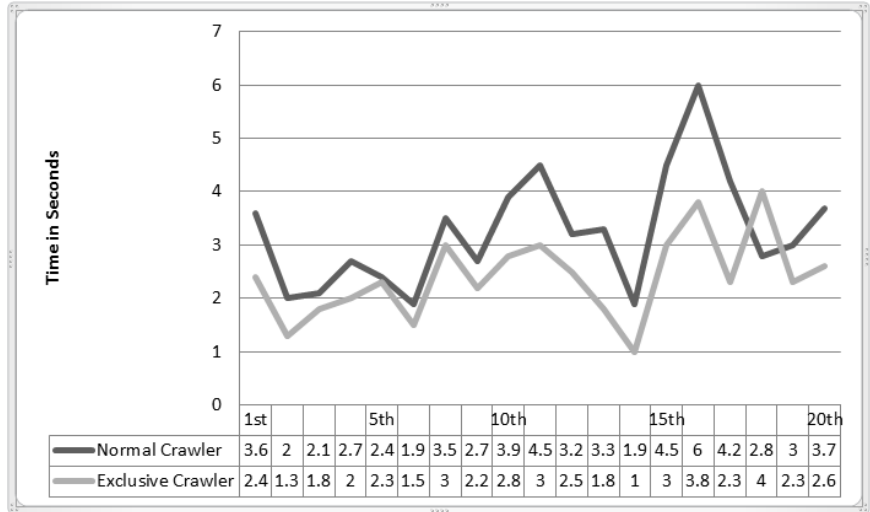

| | 1st | | | 5th | | | 10th | | | 15th | | | 20th |
|---|---|---|---|---|---|---|---|---|---|---|---|---|---|
| Normal Crawler | 3.6 | 2 | 2.1 | 2.7 | 2.4 | 1.9 | 3.5 | 2.7 | 3.9 | 4.5 | 3.2 | 3.3 | 1.9 | 4.5 | 6 | 4.2 | 2.8 | 3 | 3.7 |
| Exclusive Crawler | 2.4 | 1.3 | 1.8 | 2 | 2.3 | 1.5 | 3 | 2.2 | 2.8 | 3 | 2.5 | 1.8 | 1 | 3 | 3.8 | 2.3 | 4 | 2.3 | 2.6 |

Figure 6. comparison of crawling speed in a normal crawler and an exclusive one in identical conditions (to crawl a 20-page web site)

In the identical conditions it is obvious that the exclusive crawler is faster than the normal one. Since the exclusive crawler:

- Only searches in its own web site and lack a Follower Function to add and crawl external links.
- Lacks access checker function to check Robots.txt and authority to access is awarded by the webmaster.
- Each site's database is smaller than that of the search engine and there is no need to a complicated algorithm to avoid saving sites, pages, keywords, etc.

## 6.Conclusion

Application of exclusive web-based crawler is a proper method to sve correct and applied information in search engine database and consequently displaying data relevant to users' demand since in this method the webmaster himself gives his site's information to the search engine database. This method has some advantages including: information is up-to-date (because of database update by the webmaster as any changes occure), deleted pages are not saved and consequently no 404 error is displayed and crawling overload declines on the side of search engine servers. Using these crawlers meets search engine useres' expectations to gain better results for their queries and webmasters' expectation to see their sites in search engine results.

**Authors**

1. **Ali Tourani** is a Bachelor student at the University of  Guilan, Guilan, Iran since 2008 in the department of Software Engineering. He's also a Member of Cryptography Scientific Association of Guilan University Since 2011 . He works as member of Programming Team of Engineering Faculty Since 2009 and 'Iranian Search Engine Implementation' Project Since 2012 . He is interested in Search Engines, Cryptography Algorithms And Socket Programming. 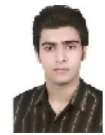

2. **Amir Seyed Danesh** is a PhD student at University of Malaya, Kuala Lumpur, Malaysia since 2008 in the department of Software Engineering. He worked as software engineer for five years in Iran. He received the M.Sc degree from Shahid Beheshti University, Tehran, Iran in Software Engineering in 2006. He is interested in Requirements Engineering, Software Release Planning, methods and Requirements Prioritization. 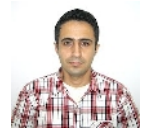